\documentclass[notitlepage,aps,epsfig,showpacs,floats,amssymb,amsmath,floatfix,groupedaddress,superscriptaddress]{revtex4-1}
\usepackage{amsfonts,amssymb,stmaryrd,latexsym,amsmath,braket}
\usepackage{graphicx,subfigure}
\usepackage{comment}
\usepackage{times}
\usepackage{slashed}
\usepackage{bm}
\usepackage{braket}
\usepackage{appendix}
\usepackage{enumitem}
\usepackage{multirow}
\usepackage{array}
\usepackage{tabularx}
\usepackage[colorlinks=true,backref=true,linktocpage=true,citecolor=blue,urlcolor=blue,linkcolor=blue,pdfpagemode=UseOutlines]{hyperref}
\usepackage[dvipsnames]{xcolor}

\begin{document}

\title{Charge Regulation Effect on Nanoparticles Interaction Mediated by Polyelectrolyte}

\author{Vijay Yadav}
\affiliation{Department of Physics, Indian Institute of Technology Jodhpur, Jodhpur, Rajasthan 342 030, India}

\author{Prabhat K. Jaiswal}
\email{prabhat.jaiswal@iitj.ac.in}
\affiliation{Department of Physics, Indian Institute of Technology Jodhpur, Jodhpur, Rajasthan 342 030, India}

\author{Rudolf Podgornik}
\email{Deceased}
\affiliation{School of Physical Sciences and Kavli Institute for Theoretical Sciences, University of Chinese Academy of Sciences, Beijing 100049, China}
\affiliation{Wenzhou Institute, University of Chinese Academy of Sciences, Wenzhou, Zhejiang 325000, China}
\affiliation{CAS Key Laboratory of Soft Matter Physics, Institute of Physics, Chinese Academy of Sciences, Beijing 100190, China}
\affiliation{Department of Physics, Faculty of Mathematics and Physics, University of Ljubljana, 1000 Ljubljana, Slovenia}

\author{Sunita Kumari}
\email{sunita@iitj.ac.in}
\affiliation{Department of Physics, Indian Institute of Technology Jodhpur, Jodhpur, Rajasthan 342 030, India}

\date{\today}

\begin{abstract}
The ability to precisely control surface charge using charged polymers is fundamental to many nanotechnology applications, enabling the design and fabrication of materials with tailored properties and functionalities. Here, we study the effect of charge regulation (CR) on the interaction between two nanoparticles (NPs) mediated by an oppositely charged polyelectrolyte (PE) in an electrolyte solution. To this end, we employ a hybrid CR Monte Carlo / molecular dynamics simulation framework to systematically explore the effects of pH, salt concentration, and polymer chain length on NP surface charge behavior. For comparison, we also conduct molecular simulations under constant charge (CC) conditions. Our results reveal that CR enhances PE adsorption onto NP surfaces compared to the CC case, where polymer bridging dominates across a wide range of NP intersurface separations. This enhanced adsorption under CR leads to a weak net repulsion driven by osmotic forces. In contrast,  the CC model yields a stronger net attraction due to the bridging force. Furthermore, we find that the CR effects are more pronounced at low salt concentration, whereas at high salt concentration, counterion screening dominates in both CR and CC cases, diminishing the CR effect. These findings highlight the importance of incorporating charge regulation in characterizing nanoparticle interactions within a complex biochemical environment, particularly in low salt concentrations.

\end{abstract}

\maketitle

\section{introduction}

Nanoparticles (NPs) exhibit remarkable properties due to their nanoscale dimensions, primarily due to their high surface area-to-volume ratio and distinct physicochemical characteristics \cite{Deirram2019,Lin2020,Ly2024}. These unique combinations make NP exceptionally suitable for loading materials and interacting with biological systems, leading to enhanced bioreactivity \cite{Deirram2019,Lin2020,Ly2024}. Furthermore, charged polymers or polyelectrolytes (PEs) offer a versatile tool for precisely tuning the physical and chemical properties of these NPs~\cite{Granfeldt1991,Muthu2023,Deirram2019,MateosMaroto2021,Ulrich2005,Ulrich2006,Carnal2011,Longo2013,Stornes2021,Lu2021}. This is crucial for various applications, especially in the medical~\cite{dela2012,Prakash2012, curtis_colloidal_2018,Deirram2019,Lin2020,Veider2023, Ly2024} and industrial fields~\cite{Karnik2005,Lu2021,Ritt2022}. Interestingly, in response to their local bathing environment, both PEs and NPs can dynamically adjust their ionizable surface groups (consequently modulating their electrostatic interaction)~\cite{Muthu2023, Yuan2022, curk_accelerated_2022,curk_charge_2021, Bian2021,Ruixuan2023,Kumari2024}. Such a charge controlling mechanism is commonly called ``charge regulation" (CR), and it underpins a wealth of remarkable yet complicated functional and structural motifs~\cite{Lund2013,Muthu2023, Yuan2022, curk_accelerated_2022,curk_charge_2021, Bian2021,Xiong2021}. Nevertheless, to take advantage of NPs in different applications, we first need a better fundamental understanding of the \textcolor{blue}{role} of CR mechanisms on NPs interactions to predict their behavior in complex PE solutions. 

 In the low salt concentration limit, the interaction forces between two like-charged species are primarily governed by strong repulsive forces. This repulsion arises from the overlap of diffusive double layers formed by counterions covering the charged surfaces~\cite{israelachvili_intermolecular_2011}.

 A single PE in a system of two or more colloids (1–10 nm) can mediate an attractive bridging interaction when the colloids carry charges opposite to that of the PE~\cite{kesson1989,Miklavic1990,podgornik_colloidal_1995,Podgornik2003}. These bridging interactions arise from the fact that some of the monomers of PE are adsorbed on both surfaces, and the rest of the monomers remain in the middle, which provides an entropy-elastic force between them~\cite{podgornik_colloidal_1995}. An attempt to describe the PE complexes on the membrane was first elucidated by Muthukumar within the framework of mean field arguments~\cite{Muthukumar1987}. Based on the Edwards–deGennes self-consistent mean-field theory (SCMF)~\cite{Edwards1965,Gennes1969} on the screened Coulomb potential interactions, Opheusden showed short-range strong attractive bridging interactions~\cite{Opheusden1988}. However, direct surface–surface contact, which is inherently repulsive, was not accounted for in the study~\cite{Opheusden1988}. 
 
 An extension of the traditional mean-field theory was developed to incorporate non-electrostatic interactions, particularly excluded volume effects between monomers~\cite{Borukhov1995,Borukhov1997,Borukhov1999,Biesheuvel2005}. The SCMF was successfully applied to planar interfaces and PE adsorption on non-planar-like charged macroions~\cite{Gurovitch1999} but failed to explain the PE-mediated interaction between these small macroions. In this particular case, another elegant formulation based on a Gaussian variational ansatz~\cite{Podgornik1993}  for the effective Hamiltonian of the free~\cite{podgornik_colloidal_1995} or grafted~\cite{Podgornik2003} PE chains was found to be more convenient~\cite{Podgornik1991,Podgornik1993}. In addition, the phenomenon of ``overcharging"~\cite{Vagharchakian2004,Dobrynin2001,Grosberg2002,Lenz2008} occurs in PE adsorption when multiple PEs are adsorbed on a surface in excess, resulting in charge reversal of the decorated surface. This can be understood through the concepts of correlated manner, where like-charged PE chains repel each other~\cite{Dobrynin2000}, and ``charge fractionalization," where the charge of a PE is not completely transferred to the surface~\cite{Nguyen2002,Nguyen200}. In the limit of higher salt concentration, the repulsive component of the interaction is partly due to directly screened electrostatic repulsions and partly due to PE brush repulsions between adsorbed monomers on charged surfaces~\cite{Dahlgren1993}. The existing various studies indicate that NP surface characteristics and conformation of PE are dramatically triggered by the properties of the PE itself (like chain length, flexibility, and charge density) and by solution pH and valency~\cite{podgornik_colloidal_1995, Podgornik2006,Ulrich2005, Ulrich2006,Carnal2011,Stornes2018}.

Numerous theoretical~\cite{andrey_wkb, andrey_wkb2} and computational works~\cite{andrey_mc, andrey_mc2} have provided valuable insights into PE adsorption on planar and curved surfaces. While advancements have been made in understanding the effect of PE on interaction between charged surfaces (steric, bridging, and depletion interactions), many studies widely adopted the constant charge (CC) conditions~\cite{kesson1989,Miklavic1990,Podgornik1991,Podgornik1993,podgornik_colloidal_1995,Podgornik2006,Szilagyi2014} or assume only one component (either PE or NPs) carries fixed charges~\cite{Netz2003, Ulrich2005,Ulrich2006,Carnal2011,MateosMaroto2021,Stornes2018,Narambuena2021}. This simplification overlooks the complex interplay of dynamic charges between both PE and NPs, which is crucial for accurate modeling and prediction of their behavior in solution. Attempts in this direction have so far been made only on PE adsorption on a single NP~\cite{Stornes2021} or flat surfaces~\cite{yuanChargeRegulationEffects2024}, suggesting that the concurrent effect of CR on both PE and charged surfaces can lead to non-trivial and unexpected interaction effects that are not captured by the traditional constant charge CC model~\cite{Stornes2021,yuanChargeRegulationEffects2024}. In essence, a more holistic approach is needed, which considers the dynamic charges (acid/base groups) occurring on both the PE and the NP.

In the 1970s, Ninham and Parsegian gave the first qualitative explanation of the CR effect, which couples the nonlinear Poisson-Boltzmann (PB) theory by incorporating variable surface charge densities governed by chemical equilibria for nonpolar planar surfaces in an aqueous solution~\cite{Ninham1971}. However, the discrete nature of the ions and surface functional groups breaks down this model~\cite{Ninham1971}, which has been shown to play a crucial role in the CR mechanism~\cite{Bakhshandeh2019,Bakhshandeh2020,Bakhsh2020,Gomez2021}. In several works, the CR formulation for surface binding sites was evaluated via the law of mass action~\cite{Chan1975, Prieve1976,vonGrnberg1999,PericetCamara2004} and, separately, by modifying the surface-site partition function (free energy) approach~\cite{Longo2012,Longo2013,Adi2016,Markovich2014,Diamant1996,Everts2016}, Frumkin–Fowler–Guggenheim isotherm model for screening effect and interaction between two macroions~\cite{Ruixuan2023,Kumari2024,Koopal2020}. The conventional PB theory, while generally accurate for modeling simple systems like planar charged surfaces or colloids near charged surfaces, faces limitations when dealing with complex systems such as high-charge-density scenarios, multivalent ions, and PE~\cite{Muthukumar2010,Blanco2019,Stornes2021}. Unfortunately, no purely theoretical approach is possible that covers a broad spectrum of biophysicochemical parameters to study NP–PE interactions.  On the other hand, computer simulation techniques are more versatile and flexible than theoretical models, since they can capture CR dependence on multiple system parameters~\cite {curk_charge_2021,curk_accelerated_2022,Yuan2022,Colla2024}. 

In this paper, we present a comprehensive analysis of how CR influences the interaction between two NPs bearing dissociable surface groups in the presence of an oppositely charged, ionizable  PE. We identify key effects of CR on the complexation behavior of the PE with the NPs in a salt solution. To model this system, we employ a recently developed hybrid charge-regulation Monte Carlo/molecular dynamics (CR-MC/MD) simulation approach~\cite{Yuan2022}.  To the best of our knowledge, this is the first study to investigate NP–NP interactions mediated by a PE where both the NPs and the PE undergo charge regulation. Although related studies have been conducted by the Ganeshan group, their approach applied charge regulation to the polymer while modeling the NPs (proteins) using fixed-charge patches of positive and negative ions~\cite{samanta_influence_2020, samanta_influence_2021}. In contrast, our model allows all components to respond dynamically to the local chemical environment, offering a more realistic representation of the system. Experimental studies have shown that such comprehensive charge regulation is critical for accurately capturing NP–PE interactions: Huang et al.~\cite{huang} observed pH-dependent adsorption and force modulation between cationic PE and silica surfaces consistent with CR, while Popa et al.~\cite{popa} measured interaction forces between latex particles coated with poly (sodium 4-styrenesulfonate) that were quantitatively described only when boundary conditions of CR were considered. By developing a more realistic understanding of such systems, this work offers critical insights for designing more effective technologies in areas ranging from biomedical applications to sustainable materials development.

The outline of the paper is as follows: The details of the model underlying the simulation methodology and force calculation between various charge components are presented in Sec.~\ref{simulation} and ~\ref{sec:force}, respectively. Next, in Sec.~\ref{result}, the analytical aspects of our results are discussed. We delineate our final remarks in the last Sec.~\ref{remarks}. 
 
\section{MODEL AND SIMULATION}
\label{simulation}

\begin{figure*}[htbp]
    \centering
    \includegraphics[width=0.9\textwidth]{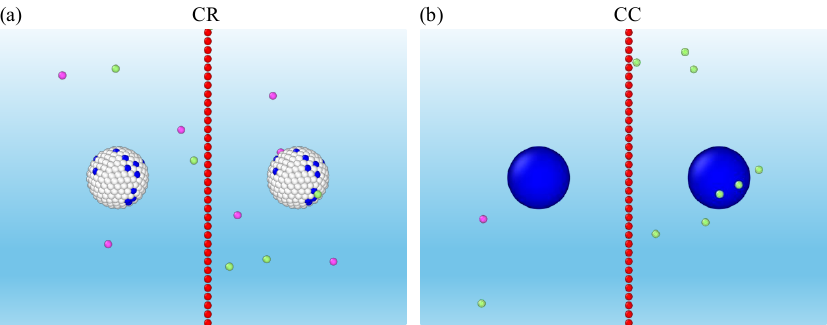}
    \caption{\label{fig:fig1} The initial configurations of our modeled systems. (a) A snapshot of the charge regulation (CR) simulation, where nanoparticles (NPs) coated with base groups ($pK_b = 5$) are separated by an intersurface distance ($D=15 \ell_B$). A polyelectrolyte (PE) chain composed of $N=40$ acid groups ($pK_a = 5$) is positioned between the NPs. Red beads represent the charged monomers of the PE, while blue and white sites correspond to the NPs charged and neutral base groups, respectively. Green and magenta beads represent positively and negatively charged ions, respectively. (b) A snapshot of the constant charge (CC) simulation approximates the CR model. Here, NPs are assigned a fixed charge equal to the average charge on NPs in the CR simulation, and this charge is placed at the center of the NPs. A PE chain of $N=40$ monomers, bearing fixed charges equal to the average monomer charge obtained in the CR simulation, is positioned between the NPs.}
\end{figure*}

We consider a system composed of two spherical NPs in the vicinity of an oppositely charged PE, as illustrated in Fig.~\ref{fig:fig1}. The system is immersed in a monovalent electrolyte solution (salt, protons, and hydroxyl ions) maintained at pH $=7$. Electrostatic interactions in the system are characterized by the Bjerrum length, $\ell_B = q^2/4\pi \varepsilon\varepsilon_0k_B T$, which represents the distance at which the electrostatic interaction between two unit charges ($q$) equals the thermal energy ($k_B T$). We use a fixed Bjerrum length of $\ell_B = 0.72$ nm in all simulations, corresponding to the dielectric properties of water at room temperature. Here, $\varepsilon$ and $\varepsilon_0$ denote the dielectric constant of the solvent and the vacuum permittivity, respectively. We simulated a cubic box with a linear size of $80\ell_B$ with periodic boundary conditions. 

A set of 256 base groups (positive sites) with dissociation constant $pK_b$ is uniformly distributed over each spherical shell ($3\ell_B$) and firmly attached to them according to the electron distribution in the Thomson problem~\cite{tp06}. The PE is modeled as a bead-spring chain of $N$ weak acid groups (negative sites) with a dissociation constant $pK_a$~\cite{Baschnagel2000, Glotzer2002, KROGER2004, Peter2009}. Here, an acid/base refers to particles that can acquire a $q=[0, \mp 1]$ charge. The NPs and PE are allowed to exchange ions with an electrolyte solution, although exhibiting overall electroneutrality. The ionizable groups undergo reactions $A^0 \rightleftharpoons A^- + H^+$ and $B^0 \rightleftharpoons B^+ + OH^-$, respectively as developed by Curk and Lujiten~\cite{curk_accelerated_2022}. For simplicity, we also assume that the monomers, surface sites, and solution mobile ions are of the same size ($2r=\ell_B$). In this method, the velocity-Verlet algorithm updates the system configurations, while CR-MC moves are employed to sample the ionization states~\cite{curk_charge_2021,curk_accelerated_2022}. 

To compare with the constant charge (CC) approximation, we assign fixed charges corresponding to the average ionization states observed in the CR simulations. Specifically, in the CC setup, each nanoparticle (NP) is modeled as a sphere with radius $4\ell_B$, bearing a constant charge equal to the average charge of an NP in the CR case. Likewise, each PE monomer is assigned a fixed charge equal to its average CR charge. The CC simulations are carried out using molecular dynamics under the same integration scheme and system parameters as the CR simulations. Counterions are included in both setups to maintain overall charge neutrality. All simulations were performed in reduced units with a time step of $0.005\tau$, using a Langevin thermostat with a damping constant of $1.0\tau$ to maintain the temperature at $T=1.0$.

We position the linear PE chain at the midpoint between the two NPs, as illustrated in Fig.~\ref{fig:fig1}. The bonded interactions between adjacent PE monomers are modeled using a harmonic potential:
\begin{equation}
U_B(r) = \frac{1}{2}K_F(r - r_0)^2~, 
\end{equation}

where $K_F$ is the harmonic force constant and $r_0$ is the equilibrium bond length. Excluded-volume interactions between all non-bonded particles are captured using an expanded Lennard-Jones (LJ) potential.

\begin{equation}
    U_{LJ}(r_{ij})=
    \begin{cases}
      4\varepsilon_{LJ} \left[ \left(\frac{\sigma}{r_{ij}-\Delta}\right) ^{12}  - \left(\frac{\sigma}{r_{ij}-\Delta}\right) ^{6} \right] , & r_{ij} \leq r_c^*~,\\
      0~, & r_{ij}>r_c^*~,
    \end{cases}
    \label{lj}
  \end{equation}
with $\Delta$ the expanded distance and $r_c^* = \Delta + 2^{1/6} \sigma $ the cutoff. The expanded LJ potential for different particles of the model system is listed in Table~\ref{tab:table1}.

Furthermore, electrostatic interactions between charged moieties are calculated by the following equation,
\begin{equation}
U_{coul} = 
\begin{cases}
\frac{{\cal C}}{\varepsilon}\frac{q_i q_j}{r_{ij}}, & \text{if}\ r_{ij} \leq r_{long}^*~,\\
      0~, & r_{ij}>r_{long}^*~,
\end{cases}
\end{equation}
where $r^*_{long}$ is the cut-off value for the long-range Coulomb interaction. The term ${\cal C} = 1/4\pi \varepsilon_0$ denotes the energy-conversion constant, where $\varepsilon$ is the dielectric constant of the solution, $q_i$ and $q_j$ are the charges on the particles, and $r_{ij}$ is the distance between them. From here, we proceed to evaluate the total force between charged NPs in a PE solution.

\begin{table}[ht]
\caption{\label{tab:table1} Lennard-Jones interaction parameters between all particle types, categorized as small (mobile charge carriers, surface sites, and monomers) with radius $r$ and large (NPs) with radius $R$. In all cases, the LJ energy is $\varepsilon_{LJ} = k_B T$ and $\sigma = \ell_B$.}

\begin{tabularx}{0.3\textwidth} { 
  | >{\centering\arraybackslash}X 
  | >{\centering\arraybackslash}X | }
 \hline
Particle type & Value of $\Delta$ \\
 \hline
$R \leftrightarrow R$ & $2R - 2r$ \\
\hline
$R \leftrightarrow r$ & $R - r$ \\
\hline
$r \leftrightarrow r$ & $0$ \\
\hline
\end{tabularx}
\end{table}

\section{Computation of total force between two nanoparticles}
\label{sec:force}
\begin{figure}[htbp]
    \centering
    \includegraphics{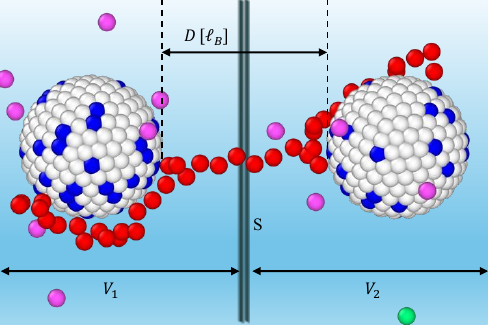}
    \caption{\label{fig:fig2} Schematic of the system for the force calculation. A fictitious plane, $S$, of width $0.25\ell_B$ is placed at the midpoint, which divides the system into two halves, $V_1$ and $V_2$. The color scheme is the same as in Fig.~\ref{fig:fig1}.}
\end{figure}

We now focus on the analytical form of the total force between two NPs due to bathing the PE solution. In the limit of CC, the interaction forces between charged colloids mediated by PEs have already been discussed in detail in Ref.~\cite{podgornik_colloidal_1995}. However, we present a pedagogical derivation of these results to ensure that the paper is self-contained. To evaluate the total force acting between two NPs, we introduce a fictitious plane, $S$, of dimension $ 0.25\ell_B \times 80\ell_B \times 80\ell_B$, positioned at the midpoint between the two NPs (see Fig.~\ref{fig:fig2}) ~\cite{podgornik_colloidal_1995, Granfeldt1991}. The system is divided into two sub-volumes, $V_1$ and $V_2$, corresponding to the left and right halves of the simulation domain, with $S$ acting as the dividing midplane. Given Ref.~\cite{podgornik_colloidal_1995}, the total force ($F_X$) acting along the radius vector joining the two NPs can be cast as,
\begin{equation}
\begin{split}
F_X &= k_{B}T \int_{(S)} d^2 \mathbf{r}\, \rho_{1}(\mathbf{r}) \\
&\quad + \int_{(V_1)} \int_{(V_2)} d^3\mathbf{r_1} \, d^3\mathbf{r_2} \, 
\rho_{2}(\mathbf{r_1},\mathbf{r_2}) \, \mathbf{f_x} (\mathbf{r_1},\mathbf{r_2})~.
\end{split}
\label{total}
\end{equation}

The first term of Eq.~\ref{total} represents the momentum exchange between $V_1$ and $V_2$ due to the movement of ions/particles across the midplane. It has the form of an ideal osmotic pressure~\cite{Lovett1992}. The second term involves the microscopic force $\mathbf{f_x} (\mathbf{r_1 ,r_2})$, which acts between a particle at position $\mathbf{r_1} \in V_1$ and another at $\mathbf{r_2} \in V_2$ projected along the direction normal to the dividing plane $S$. Here, $\rho_{1}(r)$ refers to the single particle density at the midplane and $\rho_{2}(\mathbf{r_1},\mathbf{r_2})$ is the two-particle correlation function for the system. This expression for force (Eq.~\ref{total}) is related to the contact theorem in plan-parallel geometry~\cite{Lovett1992}. The microscopic force $\mathbf{f_x} (\mathbf{r_1 ,r_2} )$ comprises several contributions: the electrostatic force between the polymer beads, the electrostatic force between the polymer beads and the two NPs, and the configurational (elastic) forces arising from bonded polymer segments whose connecting vector intersects the dividing plane $S$. As in Ref.~\cite{podgornik_colloidal_1995}, we obtain the total force,
\begin{equation}
    F_X = F_X(\text{osm}) + F_X(\text{cor}) + F_X(\text{nn}) + F_X(\text{bridge})~, 
    \label{total1}
\end{equation}
where $F_X(\text{osm})$ denotes the contribution from the osmotic force due to the momentum exchange mechanism across planes. The force component $F_{X}(\text{cor})$ is due to the correlation effect, and the contribution due to the electrostatic interaction between CR NPs is captured by $F_{X}(\text{nn})$. Since the polymer can act as a bridge between NPs when crossing mid-plane $S$, there is an additional term $F_{X}(\text{bridge})$ in the above Eq.~\ref{total1}.

The osmotic force component, $F_X(\text{osm})$, is calculated using the relation $F_X(\text{osm}) = kT<n>$, where $<n>$ is the average number of particles inside the plane $S$. The correlation force component, $F_X(\text{cor})$, is obtained by evaluating the electrostatic interactions between (i) all mobile particles in subvolumes $V_1$ and $V_2$; (ii) all mobile particles in $V_1$ and the NP in $V_2$; and (iii) the NP in $V_1$ and all the mobile particle in $V_2$. $F_X(\text{nn})$ is direct electrostatic force between two NPs. We have utilized the Particle–Particle Particle–Mesh (PPPM) concept to compute electrostatic forces with a precision of  $10^{-3}$. The bridging force, $F_X(\text{bridge})$, given by
\begin{equation}
    F_X(\text{bridge}) = -K_F\left<|X_i - X_{i+1}|\right>\times n_{br}~,
    \label{bridge}
\end{equation}
where $K_F$ is the stiffness constant (set as $100$ in Lennard-Jones units), $X_i$, $X_{i+1}$ are the $x$-coordinates (along the axis normal to the midplane $S$) of two consecutive beads on a polymer chain located on opposite sides of $S$, and $n_{br}$ denotes the average number of such bridging segments. All simulations and force calculations are performed using the LAMMPS software~\cite{LAMMPS}, except for the bridging force, which is computed using an in-house Python script developed by the authors.

\section{Results and Discussion}
\label{result}

In this section, we present numerical results based on the formulation described in the previous section. To elucidate the effects of charge regulation (CR) on the PE-NP interaction and NP-NP interaction, we compare the results of CR and constant charge (CC) simulations performed under the same conditions, as outlined in Sec.~\ref{simulation}.

\begin{figure*}[htbp]
    \centering
    \includegraphics[width=0.9\textwidth]{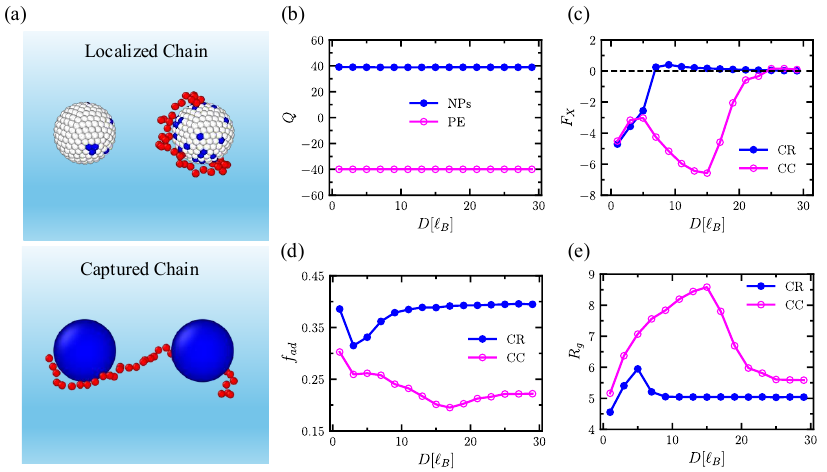}
    \caption{\label{fig:fig3}Charge regulation (CR) effects on NPs–PE interactions at a low salt concentration ($pIp = pIm = 7$) with PE bearing acidic groups ($pK_a = 5$) and NPs coated with basic groups ($pK_b = 5$). (a) Top: Snapshot of CR simulation at intersurface separation between colloids, $D = 7\ell_B$. Red and white beads denote charged and neutral acid monomers of the PE, while blue and white sites represent charged and neutral base sites. Bottom: Simulation snapshot of a constant charge (CC) simulation at the same $D$. (b) The average charge on NPs and PE as a function of $D$. (c) The total force ($F_X$) between NPs for CR and CC as a function of $D$. (d) Adsorbed monomer fraction ($f_{ad}$), within $2\sigma$ of the colloid surface as a function of $D$. (e) The radius of gyration ($R_g$) of the PE as a function of $D$. The averages are calculated over 7 (CR) and 30 (CC) equilibrated configurations.}
\end{figure*}

We begin with the case of a deionized solution, i.e., with $pIp = pIm = 7$. The corresponding system parameters are $pH = 7$, $pK_a = pK_b = 5$, and the number of monomers is fixed at $N=40$. Figure~\ref{fig:fig3}a shows simulation snapshots of CR and CC systems at a fixed intersurface separation of $D=7\ell_B$. Notably, in the CR case, the PE is adsorbed onto the surface of one of the NPs (a localized state). In contrast, under CC conditions at the same separation, PE preferentially forms a bridge between two like-charged NPs, referred to here as the captured state. Within the CC approximation, bridging interactions occur due to the electrostatic attraction between oppositely charged PE chain and the fixed surface charges of NPs, creating a strong and persistent electrostatic attraction for the oppositely charged PE segments, leading to bridging. As a result, the PE becomes partially adsorbed onto both NPs, and the intervening segment forms a bridge that generates an entropic-elastic force between them. These observations are consistent with the previous studies~\cite{Podgornik2003, Podgornik2006}. In contrast, under CR conditions, the surface charges of the NPs dynamically adjust in response to the local electrostatic environment. This flexibility allows one of the NPs to acquire a higher effective surface charge that strongly attracts the entire PE chain, resulting in its complete adsorption. This adaptive charge response suppresses polymer bridging and favors a localized state, thereby minimizing the system’s free energy through a combination of enhanced electrostatic attraction and configurational entropy gain. Further, we find that, under CR conditions, the PE is equally likely to adsorb onto either of the two NPs, reflecting the system's symmetry.

At equilibrium, the total charge on the PE and the adsorbing NPs is nearly equal in magnitude and opposite in sign for the CR parameters mentioned above, as illustrated in Fig.~\ref{fig:fig3}b. Figure~\ref{fig:fig3}c presents the total force ($F_X$) between the NPs as a function of their intersurface separation. The results show that the PE mediates an attractive interaction between like-charged NPs over a broad range of separations, consistent with previous findings for the CC case~\cite{podgornik_colloidal_1995}. In contrast, for the CR case, attractive interactions are observed only at very short intersurface separations and diminish rapidly with increasing distance, becoming weakly repulsive around $D = 7\ell_B$. This behavior results from the complete adsorption of the PE onto one of the NPs, effectively neutralizing its surface charge. As the separation increases, the dominant contribution to the total force arises from the osmotic pressure of mobile ions and monomers near the midplane, leading to a net repulsive interaction. Notably, charge regulation significantly alters the interaction landscape. As seen in Fig.~\ref{fig:fig3}c at $D = 15\ell_B$, the magnitude of $F_X$ in the CR case differs by nearly a factor of six compared to that in CC. However, at very large separations, the total forces in both CR and CC converge. This is because, in the CC scenario, the PE also eventually undergoes complete adsorption onto an NP.

A quantitative characterization of the CR effect is provided by examining the fraction of adsorbed monomers, $f_{ad}$, defined as those located within $2\sigma$ of the NP surface (see Fig.~\ref{fig:fig3}d). At very small intersurface separations, monomers remain near the NP surfaces due to limited available space, resulting in similar values of $f_{ad}$ and $F_X$ for both CR and CC simulations. However, as $D$ increases beyond $7\ell_B$, the PE in CR simulations remains fully adsorbed onto one of the NP surfaces. As a consequence, attractive interactions are suppressed because they originate from a monomer that is stretched across the midplane. In contrast, for the CC case, $f_{ad}$ decreases with increasing $D$, indicating that the PE bridges the two NPs. As separation increases, monomers span a larger distance across the midplane, leading to stronger entropic-elastic attractions. Beyond a critical separation (around $15\ell_B$), the PE in CC simulations begins to adsorb onto a single NP surface, reducing the bridging and hence diminishing the attractive force. Even at very large separations, the CR model shows nearly twice the monomer adsorption compared to CC. This suggests that CR enhances the affinity of the PE for the surface, leading to stronger adhesion. A similar suppression of bridging at high surface charge has been observed experimentally. Popa et al.~\cite{popa} used colloidal-probe atomic force microscopy to show that once the surfaces reached or exceeded charge reversal through polyelectrolyte adsorption, bridging became rare despite high polymer coverage, consistent with our CR results. This difference is also reflected in the radius of gyration, $R_g$, shown in Fig.~\ref{fig:fig3}e, which remains consistently higher in CC simulations. The increased $R_g$ indicates a more extended polymer conformation due to weaker adsorption and stronger monomer-monomer repulsions. These findings indicate that CR significantly enhances PE adsorption onto NPs, in agreement with the recent studies on PE adsorption onto planar surfaces~\cite{yuanChargeRegulationEffects2024}.

\begin{figure}[htbp]
    \centering
    \includegraphics{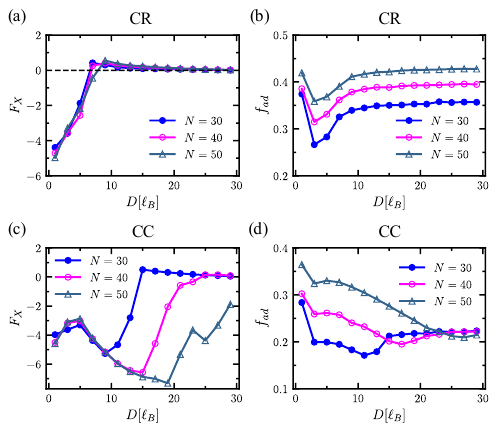}
    \caption{\label{fig:fig4} Effect of PE chain length on interaggregate force and PE adsorption across different simulation models, under the same conditions as in Fig.~\ref{fig:fig3}. (a) The total force ($F_X$) between NPs in the CR simulation. (b) Adsorbed monomer fraction onto NPs in CR simulations. (c) $F_X$ between NPs in CC simulations. (d) Adsorbed monomer fraction onto NPs in CC simulation.}
\end{figure}

Next, we examine the influence of the length of the PE chain ($N$) on the adsorption behavior and the interaggregate force ($F_X$). Simulations were performed for $N = 30$, $40$, and $50$ while keeping all other parameters identical to those used in Fig.~\ref{fig:fig3}. As shown in Fig.~\ref{fig:fig4}a, the length of the PE chain has a minimal impact on $F_X$ in CR simulations. This insensitivity can be attributed to the strong adsorption of the PE onto the surface of a single NP, as illustrated in Fig.~\ref{fig:fig4}b, which effectively neutralizes the NPs charge and suppresses long-range electrostatic interactions. Interestingly, Fig.\ref{fig:fig4}b shows that PE adsorption onto the NP surface monotonically increases with polymer chain length. This behavior is in stark contrast to previous observations for planar surfaces~\cite{yuanChargeRegulationEffects2024}, where adsorption tends to decrease with increasing chain length due to the entropic penalty associated with flattening long chains onto a surface. In our case, the curved geometry of the NP reduces the loss of conformational entropy during adsorption, allowing longer chains to maintain configurational flexibility while still forming multiple contacts with the surface. Additionally, the enhanced total charge of longer PEs increases their electrostatic attraction to the charged NP, further promoting adsorption. The CC model reveals a pronounced dependence of $F_X$ on the PE chain length. As depicted in Fig.~\ref{fig:fig4}c, longer chains lead to stronger attractive interactions at larger intersurface separations. This enhancement is driven by more effective bridging between the NPs, consistent with the observed decrease in PE adsorption onto individual NP surfaces (see Fig.~\ref{fig:fig4}d). Thus, while the interaggregate force in the CR case remains largely unaffected by the length of the PE because of complete adsorption, in the CC case the length dependence is strong because of persistent bridging interactions.

Now, we investigate the effect of the acid and base dissociation constants of the PE and NPs on the fraction of adsorbed monomers, $f_{ad}$. To this end, we systematically vary the values of $pK_a$ and $pK_b$ while keeping the salt concentrations fixed ($pIp = pIm = 7$) and maintaining a constant pH of 7. To capture the combined influence of these parameters, we define the quantity $\Delta pK = \Delta pK_a + \Delta pK_b = (pK_a - pH) + (pK_b - pH) = pK_a + pK_b - 2pH$. A positive $\Delta pK$ corresponds to a regime in which a few surface sites on the NPs are protonated and fewer monomer units on the PE are deprotonated. Consequently, both the NPs and the PE bear reduced net charges, resulting in weaker electrostatic interactions and reduced adsorption. In contrast, a negative $\Delta pK$ indicates stronger acid/base dissociation, leading to more highly charged NPs and PE chains and, hence, stronger electrostatic attraction and greater adsorption. When $\Delta pK = 0$, the system is effectively charge-balanced, representing an intermediate regime of interaction strength.

\begin{figure}[htbp]
    \centering
    \includegraphics{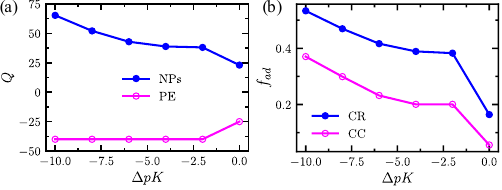}
    \caption{\label{fig:fig5} Effect of net acid and base dissociation constants on adsorbed monomer fraction ($f_{ad}$) at salt concentration $pIp = pIm = 7$ and $pH=7$ (a) Total charge on macromolecules and (b) $f_{ad}$ as a function net ionization parameter $\Delta pK = pK_a + pK_b - 2pH$. }
\end{figure}

Figure~\ref{fig:fig5}a exhibits that at $\Delta pK = 0$, the net charges on both NPs and PE are balanced and relatively low, leading to weak electrostatic interactions. As a result, the corresponding fraction of adsorbed monomers, $f_{ad}$, remains low in both CR and CC simulations, with only a minor difference between the two cases as shown in Fig.~\ref{fig:fig5}b. As $\Delta pK$ becomes increasingly negative, the net charges on the NPs and PE grow, thereby strengthening the electrostatic attraction and enhancing adsorption. However, because the PE chain length is fixed at $N = 40$, the total charge on the PE saturates beyond a certain threshold. Interestingly, the disparity between CR and CC simulations becomes most pronounced at intermediate $\Delta pK$ values. This can be attributed to the enhanced responsiveness of surface charge regulation under moderately strong electrostatic conditions. At a very negative $\Delta pK$, although the electrostatic attraction is strong, the high adsorption in both cases narrows the difference between CR and CC again. Thus, we conclude that the distinction between CR and CC is minimal at $\Delta pK = 0$, increases at intermediate values, and diminishes at large negative $\Delta pK$, where adsorption is near saturation in both regimes.

\begin{figure}[htbp]
    \centering
    \includegraphics{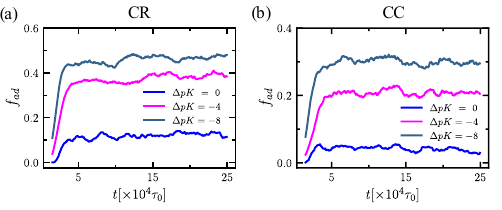}
    \caption{\label{fig:fig6} Temporal evolution of the adsorbed monomer fraction ($f_{ad}$) in (a) CR and (b) CC simulations. Blue, magenta, and metallic blue solid lines correspond to $\Delta pK = 0$, $\Delta pK = -4$, and $\Delta pK = -8$, respectively.}
\end{figure}

Furthermore, we investigate the temporal evolution of the adsorbed monomer fraction, $f_{ad}$, at various $\Delta pK$ values for both CR and CC simulations. As shown in Fig.~\ref{fig:fig6}a, CR simulations exhibit a monotonic increase in the equilibrium value of $f_{ad}$ with an increasing negative $\Delta pK$, reflecting stronger electrostatic interactions and improved PE adsorption. A similar trend is observed in CC simulations as evident in Fig.~\ref{fig:fig6}b; however, the kinetics of adsorption differ markedly. Notably, the adsorption process proceeds significantly faster in the CR simulations, reaching equilibrium values approximately twice as quickly as in the CC case. This accelerated adsorption under charge-regulating conditions highlights the dynamic adaptability of surface charge, which facilitates rapid electrostatic accommodation and polymer attachment.

\begin{figure*}[!htbp]
    \centering
    \includegraphics[width=0.9\textwidth]{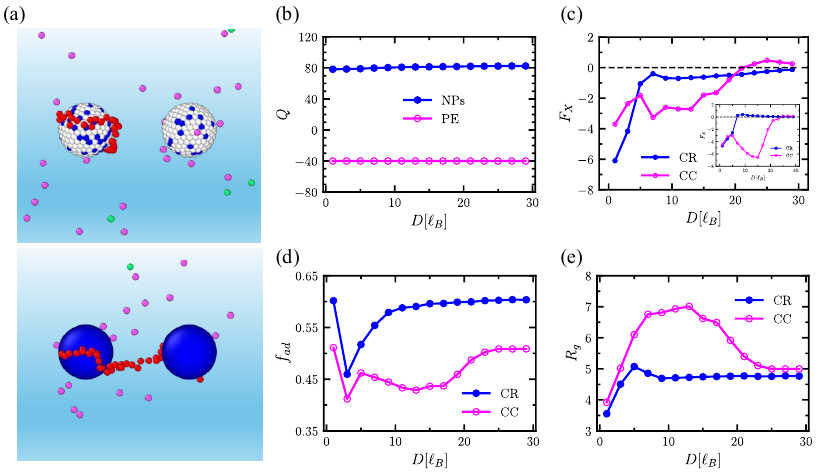}
    \caption{\label{fig:fig7} Charge regulation (CR) effects on NPs–PE interactions at low higher concentrations ($pIp = pIm = 3$) with keeping all other parameters same as in Fig.~\ref{fig:fig3}. (a) Top: Simulation snapshot of CR simulation at intersurface separation ($D$) between colloids, $7\ell_b$. The color scheme is the same as in Fig. 1. Bottom: Simulation snapshot of a constant charge (CC) simulation at the same $D$. (b) The average charge on NPs and PE as a function of $D$. (c) The total force ($F_X$) between NPs for CR and CC as a function of $D$. The insert shows $F_X$ at low salt concentration. (d) Adsorbed monomer fraction ($f_{ad}$), as a function of $D$. (e) The radius of gyration ($R_g$) of the PE versus $D$. The averages are calculated over 5 (CR) and 50 (CC) equilibrated configurations.}
\end{figure*}

Finally, we investigate the electrostatic interaction between PE and NPs at a higher salt concentration ($pIp = pIm = 3$), while keeping all other parameters identical to those used in Fig.~\ref{fig:fig3}. Simulation snapshots at $D=7\ell_b$ are depicted in Fig.~\ref{fig:fig7}a, revealing similar structural features to those observed under low salt conditions (see Fig.~\ref{fig:fig3}a). However, as illustrated in Fig.~\ref{fig:fig7}b, the charge on the NPs is nearly twice the magnitude and opposite in sign to that of the PE, due to the increased availability of free ions participating in surface ionization reactions. 

The impact of salt concentration on the interaggregate force is depicted in Fig.~\ref{fig:fig7}c. In the CC simulations, polymer bridging persists, but at a shorter range of separation than in the low-salt scenario, which is consistent with previous findings~\cite{podgornik_colloidal_1995}. In contrast, the CR simulations show no qualitative change in the force profile with increasing salt concentration, except for a noticeable downward shift in $F_X$. This reduction in net force arises from the increased population of counterions (primarily anions) required to neutralize the higher positive surface charge on the NPs. Overall, these observations suggest that, under CC conditions, NP-NP interactions are strongly PE-mediated at low salt concentrations but become progressively screened as the ionic strength increases. In contrast, under CR conditions, the interaggregate force remains dominated by counterions across salt concentrations, reflecting the dynamic charge regulation and strong ion-mediated screening intrinsic to these systems.
Furthermore, we observe a significant increase in PE adsorption for both CR and CC simulations at high salt concentrations, as shown in Fig.~\ref{fig:fig7}d. Importantly, the difference in adsorption between CR and CC systems is significantly reduced compared to that seen under low-salt conditions (see Fig.~\ref{fig:fig3}d), which can be attributed to the higher surface charge on the NPs under these conditions. A comparable trend has also been observed experimentally. Huang et al.~\cite{huang} showed that increasing salt concentration shortened the range of interaction forces and reduced distinct charge-regulation effects, although adsorption remained slightly higher than in the constant charge, in agreement with our findings. The adsorption difference is mainly due to the enhanced surface affinity in the CR model, which persists even under strong ionic screening. A similar convergence is evident in the radius of gyration (see Fig.~\ref{fig:fig7}e), where the disparity between CR and CC simulations diminishes with increasing salt concentration, indicating more compact PE conformations in both cases. 

These results highlight that although the CR and CC systems exhibit clear differences at intermediate separations, especially under low ionic strength, these differences diminish at larger separations and higher salt concentrations. In CR simulations, increased salt primarily leads to a suppression of interaggregate forces due to enhanced screening by counterions. In contrast, CC simulations remain sensitive to salt, with polymer-mediated bridging interactions being progressively diminished. Nevertheless, in both scenarios, higher salt promotes greater PE adsorption and a reduction in $R_g$, reflecting more compact and surface-associated polymer configurations.

\section{Conclusion}
\label{remarks}

The highly versatile nature of charged polymers or polyelectrolytes (PEs) provides enormous opportunities for controlled manipulation of nanoparticles (NPs) physicochemical properties. The adsorption of PE on the surfaces of NPs plays a critical role in modulating their stability and aggregation behavior in electrolyte solution. This is crucial for various applications. In this study, we have systematically explored the influence of charge regulation (CR) on the interactions between two NPs coated with base groups in the presence of an oppositely charged single weak PE in a monovalent solution. We employed a hybrid charge-regulation Monte Carlo/molecular dynamics (CR-MC/MD) framework to investigate PE adsorption and its impact on interparticle forces. In addition, constant-charge (CC) molecular dynamics simulations were carried out to assess the impact of charge regulation.

We find that at low salt concentrations ($pIp = pIm = 7$), CR significantly enhances PE adsorption onto NP surfaces, leading to the collapse of the PE chain on one of the NPs. This adsorption strongly suppresses electrostatic repulsion by effectively neutralizing the NP surface charges, resulting in weak net interparticle interactions across a wide range of NPs separations. In contrast, the CC simulation shows long-range attractions driven by polymer bridging, with the PE chains adopting extended conformations connecting both NPs. 

Additionally, the dependence of intermolecular forces on PE chain length was found to be markedly different between these two models. In CC simulations, longer PE chains enhance bridging and hence attractive interactions. In CR simulations, the behavior is largely chain-length independent due to saturation of adsorption sites. 

We further investigated the role of acid-base dissociation constants by tuning the net ionization parameter, $\Delta pK = pK_a + pK_b - 2pH$. This parameter captures the joint contribution of the dissociable groups on both the PE and the NP surfaces. The results demonstrate that the difference between CR and CC is most significant at intermediate $\Delta pK$ values, where partial ionization allows CR to dynamically respond to local electrostatic environments, leading to stronger adsorption and reduced interparticle forces. Notably, CR consistently exhibits faster and more complete adsorption dynamics, emphasizing its responsiveness and kinetic favorability over the CC approximation.

Finally, we show that at high salt concentration ($pIp = pIm = 3$), electrostatic screening reduces the effective interactions in both CR and CC simulations. Under these conditions, the differences in the extent of adsorption and radius of gyration between the two models diminish. However, even in the high-salt regime, charge regulation continues to exhibit greater PE adsorption, although the difference in the inter-particle forces becomes less pronounced. This underlines the ion-sensitive nature of charge regulation, where the balance between surface ionization and ionic screening governs both adsorption and colloidal/NPs stability. Our findings, in line with experimental observations, provide strong validation for our CR-MC/MD framework. These results further indicate that surface-tunable charge regulation can be used as a strategic tool for stabilizing nanoparticle suspensions under physiological ionic conditions. Future work could explore active tuning of CR by external stimuli such as electric fields or pH gradients.

\section{Acknowledgment}
P.K.J. acknowledges the financial support from the Anusandhan National Research Foundation, India (Grant Nos. CRG/2022/006365 and ITS/2024/001866) and IIT Jodhpur for a Seed Grant (No. I/SEED/PKJ/20220016). SK acknowledges the financial support from the Anusandhan National Research Foundation, India (grant No. EEQ/2023/000676) and IIT Jodhpur for a research initiation grant (grant no. I/RIG/SNT/20240068).

\bibliographystyle{apsrev4-1}
\bibliography{PE_NP}

\end{document}